\documentclass[twocolumn,aps,prd,amsmath,amssymb,floatfix,showpacs]{revtex4}

\usepackage{bm}
\usepackage{amsmath}
\usepackage{epsf}

\begin{document}

\title{Time variation of a fundamental dimensionless constant}

\author{Robert J. Scherrer}
\address{Department of Physics and Astronomy, Vanderbilt University,
Nashville, TN ~~37235}

\begin{abstract}
We examine the time variation of a previously-uninvestigated fundamental dimensionless
constant.  Constraints are placed on this time variation using historical
measurements.  A model is presented for the time variation,
and it is shown to lead to an accelerated expansion for the universe.
Directions for
future research are discussed.
\end{abstract}
\maketitle

\section{Introduction}
Physicists have long speculated that the fundamental constants
might not, in fact, be constant, but instead might vary with time.
Dirac was the first to suggest this possibility \cite{dirac}, and 
time variation of the fundamental constants has been investigated numerous
times since then.
Among the various possibilities, the fine structure constant
and the gravitational constant have
received the greatest attention, but work has also been done, for example, on
constants related
to the weak and strong interactions, the electron-proton mass ratio,
and several others.

It is well-known that only time variation of dimensionless fundamental
constants has
any physical meaning.  Here we consider the time variation of a dimensionless
constant not previously discussed in the literature:  $\pi$.  It is impossible to
overstate the significance of this constant.  Indeed,
nearly every paper in astrophysics makes use of it.  (For a randomly-selected
collection of such papers, see Refs. \cite{scherrer1,scherrer2,scherrer3,
scherrer4,scherrer5,scherrer6,scherrer7,
scherrer8,scherrer9}).

\begin{table}
\begin{center}
\begin{tabular}{|r|r|r|r|r|} \hline
Location & Time & $\pi(t)$ \\ \hline \hline
Babylon & 1900 BC & 3.125 \\ \hline
India & 900 BC & 3.139 \\ \hline
China & 263 AD & 3.14 \\ \hline
China & 500 AD & 3.1415926 \\ \hline
India & 1400 AD & 3.14159265359 \\ \hline
\end{tabular}
\end{center}
\caption{The value of $\pi$ measured at the indicated location at the indicated
time.}
\end{table}

In the next section, we discuss the observational evidence for
the time variation of $\pi$.
In Sec. III, we present a theoretical model, based on string theory,
which produces such a time variation, and we
show that this model leads naturally to an accelerated expansion
for the universe.  The Oklo reactor is discussed in Sec. IV,
and directions for future research are presented in Sec. V.

\section{Evidence for time variation of $\pi$}

The value of $\pi$ has been measured in various locations over the past
4000 years.  In Table 1, we compile a list of representative historical
measurements \cite{wkp}.
We see evidence for both spatial and time variation of $\pi$.
We will leave the former for a later investigation, and concentrate on
the latter.
In Fig. 1, we provide a graph illustrating
the time variation more clearly.
\begin{figure}
\centerline{\epsfxsize=3.8truein\epsfbox{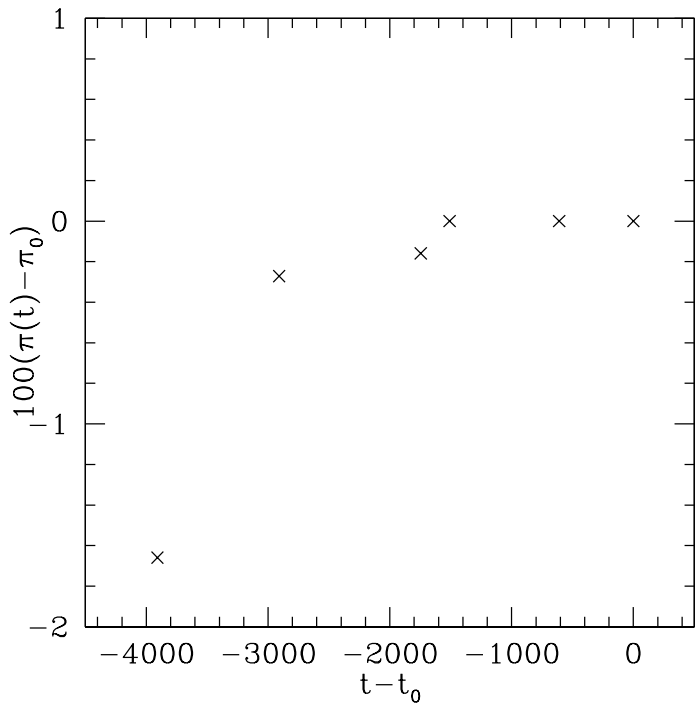}}
\caption{The value of $\pi$ relative to its present-day value, $\pi_0$, as a function
of $t-t_0$, where $t_0$ is the present time.  Time is measured in years,
and the quantity plotted on the vertical axis has been chosen to make the
time variation appear larger than it really is.}
\end{figure}
\begin{figure}
\centerline{\epsfxsize=3.8truein\epsfbox{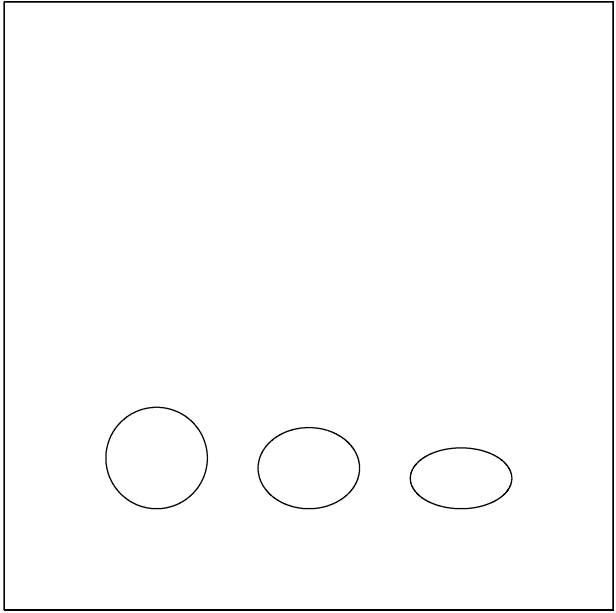}}
\caption{Geometric distortion produced by the leakage of $\pi$ into
a higher dimension.}
\end{figure}
The values of $\pi(t)$ show a systematic trend, varying monotonically
with time and converging to the present-day measured value.  The evidence for
time variation of $\pi$ is overwhelming.

\section{A theoretical model}

Inspired by string theory \cite{awi}, we propose the following model for the
time variation of $\pi$.  Consider the possibility that our observable universe
is actually a 4-dimensional brane embedded in a 5-dimensional bulk.  In this
case, ``slices" of $\pi$ can leak into the higher dimension, resulting in a
value of $\pi$ that decreases with time.  This leakage into a higher dimension
results in a characteristic geometric distortion, illustrated in Fig. 2.
Such ``leakage" has been observed previously in both automobile and bicycle
tires.  However, it is clear that more controlled experiments are necessary to
verify this effect.

It might appear that the observational data quoted in the previous section
suggest a value of $\pi$ that increases with time, rather than decreasing as
our model indicates.  Since our theoretical model is clearly correct, this
must be attributed to 4000 years of systematic errors.

Now consider the cosmological consequences of this time variation in $\pi$.
The Friedmann equation gives \cite{Fried}
\begin{equation}
\frac{\dot a}{a} = \sqrt{\frac{8\pi G \rho}{3}},
\end{equation}
where $a$ is the scale factor and $\rho$ is the total density.  At late
times $\rho$ is dominated by matter, so that
$\rho \propto a^{-3}$.
Hence,
if $\pi$ increases faster than $a$, the result will be
an accelerated expansion.  Of course, our model gives the opposite sign for the
time-variation of $\pi$, but this is a minor glitch which is probably easy
to fix.

This model for the time variation of $\pi$ has several other consequences.  It
provides a model for the dark matter \cite{DM}, and it can be used to derive
a solution to the cosmological constant coincidence problem \cite{CC}.
Further, it can be developed into a quantum theory of gravity \cite{QG}.
\\
\section{The Oklo reactor}

No discussion of the time-variation of fundamental constants would be complete
without a mention of the Oklo natural fission reactor.

\section{Directions for future investigation}

This investigation clearly opens up an entirely new direction in the study of
the time variation of fundamental constants.  The next obvious possibility
is the investigation of the time variation of $e$.  Following this, there is
a plethora of other constants that could be examined:  the Euler-Mascheroni
constant $\gamma$, the golden ratio $\phi$, Soldner's constant, and Catelan's
constant.

More speculatively, one might consider the possibility that the values of the
integers could vary with time, a result suggested by several early Fortran
simulations.  This possibility would have obvious implications for finance and
accounting.

\acknowledgments

A number of colleagues were kind enough to comment on the manuscript.  For some
reason they did not want me to use their names, so I will identify them
by their initials:  S. Dodelson, A.L. Melott, D.N. Spergel, and T. J. Weiler.


\begin{thebibliography}{99}
\bibitem{wkp} I could have checked the original references, but it
was easier to look everything up on Wikipedia.
\end{thebibliography}
\end{document}